\documentclass[aps,reprint,twocolumn,superscriptaddress,citeautoscript,showpacs,amsmath,amssymb,floatfix,prb]{revtex4-1}

\usepackage{amsmath}
\usepackage{amssymb}
\usepackage{graphicx}
\usepackage{times}
\usepackage{bm}
\usepackage{color}
\usepackage{gensymb}
\usepackage{dcolumn}

\newcolumntype{d}[1]{D{.}{.}{#1}}

\begin{document}
\title{Suppression of magnetic order in CaCo$_{1.86}$As$_{2}$ with Fe substitution:  Magnetization, neutron diffraction, and x-ray diffraction studies of Ca(Co$_{1-x}$Fe$_{x}$)$_{y}$As$_{2}$}

\author{W. T. Jayasekara}
\affiliation{Ames Laboratory, U.S. DOE, Iowa State University, Ames, Iowa 50011, USA}
\affiliation{Department of Physics and Astronomy, Iowa State University, Ames, Iowa 50011, USA}

\author{Abhishek Pandey}
\altaffiliation[Present Address:  ]{Department of Physics and Astronomy, Texas A\&M University, College Station, TX 77840-4242, USA}
\affiliation{Ames Laboratory, U.S. DOE, Iowa State University, Ames, Iowa 50011, USA}
\affiliation{Department of Physics and Astronomy, Iowa State University, Ames, Iowa 50011, USA}

\author{A. Kreyssig}
\affiliation{Ames Laboratory, U.S. DOE, Iowa State University, Ames, Iowa 50011, USA}
\affiliation{Department of Physics and Astronomy, Iowa State University, Ames, Iowa 50011, USA}

\author{N. S. Sangeetha}
\affiliation{Ames Laboratory, U.S. DOE, Iowa State University, Ames, Iowa 50011, USA}
\affiliation{Department of Physics and Astronomy, Iowa State University, Ames, Iowa 50011, USA}

\author{A. Sapkota}
\affiliation{Ames Laboratory, U.S. DOE, Iowa State University, Ames, Iowa 50011, USA}
\affiliation{Department of Physics and Astronomy, Iowa State University, Ames, Iowa 50011, USA}

\author{K. Kothapalli}
\affiliation{Ames Laboratory, U.S. DOE, Iowa State University, Ames, Iowa 50011, USA}
\affiliation{Department of Physics and Astronomy, Iowa State University, Ames, Iowa 50011, USA}

\author{\mbox{V. K. Anand}}
\altaffiliation[Present Address:  ]{Helmholtz-Zentrum Berlin f{\"{u}}r Materialien und Energie GmbH, Hahn-Meitner Platz 1, D-14109 Berlin, Germany}
\affiliation{Ames Laboratory, U.S. DOE, Iowa State University, Ames, Iowa 50011, USA}
\affiliation{Department of Physics and Astronomy, Iowa State University, Ames, Iowa 50011, USA}

\author{W. Tian}
\affiliation{Quantum Condensed Matter Division, Oak Ridge National Laboratory, Oak Ridge, Tennessee 37831, USA}

\author{D. Vaknin}
\affiliation{Ames Laboratory, U.S. DOE, Iowa State University, Ames, Iowa 50011, USA}
\affiliation{Department of Physics and Astronomy, Iowa State University, Ames, Iowa 50011, USA}

\author{D. C. Johnston}
\affiliation{Ames Laboratory, U.S. DOE, Iowa State University, Ames, Iowa 50011, USA}
\affiliation{Department of Physics and Astronomy, Iowa State University, Ames, Iowa 50011, USA}

\author{R. J. McQueeney}
\affiliation{Ames Laboratory, U.S. DOE, Iowa State University, Ames, Iowa 50011, USA}
\affiliation{Department of Physics and Astronomy, Iowa State University, Ames, Iowa 50011, USA}

\author{A. I. Goldman}
\affiliation{Ames Laboratory, U.S. DOE, Iowa State University, Ames, Iowa 50011, USA}
\affiliation{Department of Physics and Astronomy, Iowa State University, Ames, Iowa 50011, USA}

\author{B. G. Ueland}
\email{bgueland@ameslab.gov, bgueland@gmail.com}
\affiliation{Ames Laboratory, U.S. DOE, Iowa State University, Ames, Iowa 50011, USA}
\affiliation{Department of Physics and Astronomy, Iowa State University, Ames, Iowa 50011, USA}

\date{\today}

\begin{abstract}
Magnetization, neutron diffraction, and high-energy x-ray diffraction results for Sn-flux grown single-crystal samples of Ca(Co$_{1-x}$Fe$_{x}$)$_{y}$As$_{2}$, $0\leq x\leq1$, $1.86\leq y \leq 2$, are presented and reveal that A-type antiferromagnetic order, with ordered moments lying along the $c$ axis,  persists for $x\lesssim0.12(1)$. The antiferromagnetic order is smoothly suppressed with increasing $x$, with both the ordered moment and N\'{e}el temperature linearly decreasing.  Stripe-type antiferromagnetic order does not occur for $x\leq0.25$, nor does ferromagnetic order for $x$ up to at least $x=0.104$, and a smooth crossover from the collapsed-tetragonal (cT) phase of  CaCo$_{1.86}$As$_{2}$  to the tetragonal (T) phase of CaFe$_{2}$As$_{2}$ occurs.  These results suggest that hole doping CaCo$_{1.86}$As$_{2}$ has a less dramatic effect on the magnetism and structure than steric effects due to substituting Sr for Ca. 
\end{abstract}

\maketitle

\section{INTRODUCTION}
Certain compounds with the ThCr$_{2}$Si$_{2}$-type body-centered-tetragonal structure are ideal systems for studying the interplay between structural, electronic, and magnetic degrees of freedom, and the occurrence of superconductivity \cite{Johnston_2010, Stewart_2011}. The parent compounds of the $122$-Fe-pnictide superconductors [$A$Fe$_{2}$As$_{2}$ ($A=$ Ca, Sr, Ba)] are prominent examples, which, upon cooling, undergo coupled structural and magnetic phase transitions from a paramagnetic tetragonal phase to an antiferromagnetic (AFM) phase with collinear stripe-type magnetic order and an orthorhombic lattice \cite{Canfield_2010, Lynn_2009, Paglione_2010}.  The stripe-type AFM order is itinerant, with an ordered Fe moment of $\mu<1~\mu_{\textrm{B}}$ and an AFM propagation vector  $\bm{\tau}_{\textrm{\textbf{st}}}=(\frac{1}{2}~\frac{1}{2}~1)$, which angle-resolved photoemission spectroscopy and band structure calculations show is consistent with nesting between hole- and electron-type Fermi-surface pockets \cite{Johnston_2010,Stewart_2011, Canfield_2010, Lynn_2009, Paglione_2010}.  Studies on compounds in which Fe is substituted by other transition metals, such as Co \cite{Sefat_2008,Ni_2008}, Ni \cite{Li_2009}, or Cu \cite{Canfield_2009}, show that the structural and magnetic phase transitions become suppressed at high enough levels of doping, and that superconductivity occurs over a limited range of doping.

\begin{figure}[]
	\centering
	\includegraphics[width=1.0\linewidth]{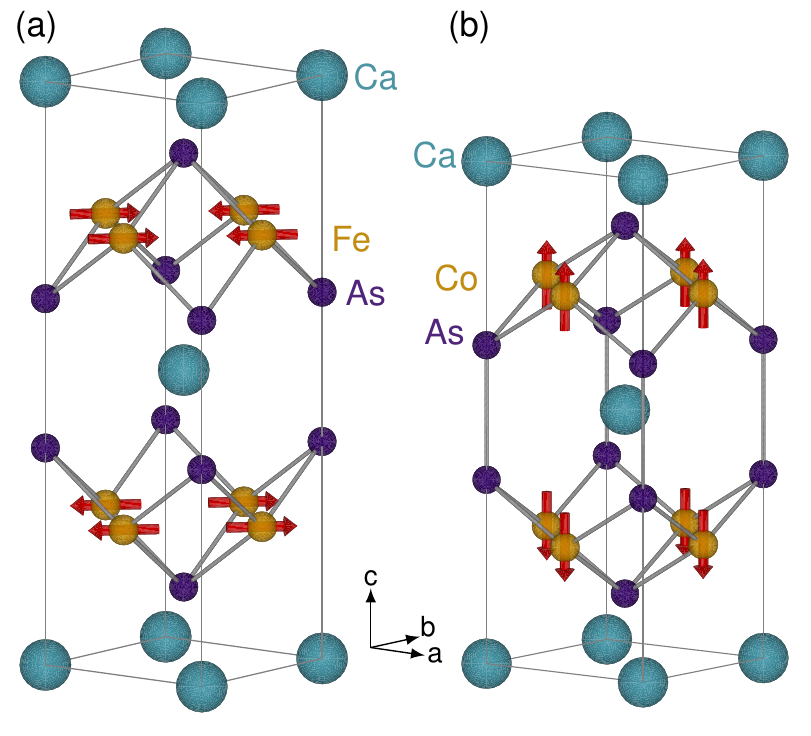}
	\caption{(Color online)  \label{Fig1} (a) The stripe-type magnetic order present in CaFe$_{2}$As$_{2}$, and  (b)  the A-type magnetic order present in CaCo$_{1.86}$As$_{2}$.  Both compounds are shown with their ambient temperature and pressure tetragonal chemical unit cells.  The diagrams were created with \textsc{vesta}\textrm{.} \cite{Momma_2011} 	}
\end{figure}

The parent $122$-compound CaFe$_{2}$As$_{2}$ undergoes a first-order transition from its paramagnetic tetragonal phase to its stripe-type AFM orthorhombic phase at a N\'{e}el temperature of $T_{\textrm{N}}=172.5$~K [see Fig.~\ref{Fig1}(a)] \cite{Goldman_2008}.  Data for samples synthesized using Sn flux show that $5.4$ to $7.5\%$ Co substitution for Fe suppresses the magnetic and structural transitions, $3.1$ to $5\%$ Co substitution causes superconductivity to first occur, and $9$ to $20\%$ Co substitution subsequently suppresses superconductivity  \cite{Harnagea_2011,Hu_2012}.  The variation in values for the level of Co substitution necessary for the various phase transitions is likely due to uncertainty in the exact amount of Co in the compounds \cite{Hu_2012}, and the various amounts of strain induced by different synthesis procedures \cite{Ran_2012,Budko_2013,Ran_2014}.
 %

Under modest applied pressure ($p_{\textrm{c}}=0.24$~GPa at $T=50$~K), CaFe$_{2}$As$_{2}$ undergoes a first-order structural phase transition from its ambient-pressure tetragonal (T) or orthorhombic phase to a collapsed-tetragonal (cT) phase characterized by an $11\%$ reduction in the ratio of its lattice parameters $c/a$, a $9.5\%$ reduction in $c$, and a $5\%$ decrease in its unit-cell volume $V_{\textrm{cell}}$ \cite{Kreyssig_2008, Goldman_2009}.  In addition, the transition quenches the Fe magnetic moment, and data show no evidence for magnetic order or AFM spin fluctuations occurring in the cT phase \cite{Pratt_2009, Soh_2013}.  The quenching of the Fe moment is understood in terms of the effective valence of Fe changing from Fe$^{+2}$ to Fe$^{+1}$ due to As-As bonds present in the cT phase \cite{Anand_2014, Quirinale_2013,Hoffman_1985, Reehuis_1990, Reehuis_1998, Anand_2012}.

CaCo$_{1.86}$As$_{2}$ is the complementary version of CaFe$_{2}$As$_{2}$ in which Fe is replaced by Co.  The compound has the same tetragonal ThCr$_{2}$Si$_{2}$ structure as CaFe$_{2}$As$_{2}$, but exists away from the ideal $122$ stoichiometry due to a $7(1)\%$ vacancy of the Co sites \cite{Quirinale_2013,Anand_2014}.  The ratio of its lattice parameters is $c/a=2.59$ \cite{Pfisterer_1980,Pfisterer_1983}, which is comparable to the value for CaFe$_{2}$As$_{2}$ in the cT phase, \cite{Kreyssig_2008} and indicates that CaCo$_{1.86}$As$_{2}$ exists in the cT phase at ambient pressure \cite{Quirinale_2013, Anand_2014}.  Data from neutron diffraction experiments on Sn-flux grown single-crystal samples show that the compound has A-type collinear itinerant AFM order below $T_{\textrm{N}}=52(1)$~K  with the ordered moments lying along the $c$ axis \cite{Quirinale_2013}.  On the other hand, thermodynamic and resistance data for CoAs-flux grown  samples indicate that the AFM phase transition occurs at either $T_{\textrm{N}}=76$~K \cite{Cheng_2012} or $\approx70$~K \cite{Ying_2012}.  Figure~\ref{Fig1}(b) shows the A-type AFM structure, which consists of Co spins ordered ferromagnetically (FM) within the $ab$ plane and aligned AFM along $c$.  Magnetization results estimate that the size of the ordered moment in Sn-flux grown samples is  $\approx0.3~\mu_{\textrm{B}}/$Co \cite{Anand_2014}, which is consistent with previous neutron diffraction results that place an upper limit of $\approx0.6~\mu_{\textrm{B}}/$Co on the ordered moment \cite{Quirinale_2013}.

Based on the cT structure of CaCo$_{1.86}$As$_{2}$, and the presence of magnetic Co$^{+1}$, it is, perhaps, not surprising that the compound has magnetic order at low temperature \cite{Anand_2014}.  On the other hand, both BaCo$_{2}$As$_{2}$ and SrCo$_{2}$As$_{2}$ have T structures at ambient pressure and do not magnetically order down to at least $T=1.8$~K.  Rather, their magnetic susceptibilities have been described as Stoner-enhanced paramagnetism lying proximate to a quantum-critical point \cite{Sefat_2009,Pandey_2013}, however, studies of K-doped BaCo$_{2}$As$_{2}$ do not support this scenario \cite{Anand_2014b}.  Additionally, inelastic neutron scattering data for SrCo$_{2}$As$_{2}$ show the presence of AFM spin fluctuations peaked at wavevectors corresponding to the stripe-type AFM order present in $A$Fe$_{2}$As$_{2}$ ($A=$ Ca, Sr, Ba)  \cite{Jayasekara_2013}, and nuclear magnetic resonance experiments find that AFM and FM spin correlations coexist in the compound \cite{Wiecki_2015}.  Thus, it is prudent to study how the lattice and magnetism change in CaCo$_{1.86}$As$_{2}$ as either Fe is substituted for Co, or Sr/Ba is substituted for Ca.

Here, we report results from magnetization, neutron diffraction, and x-ray diffraction experiments on the series of compounds Ca(Co$_{1-x}$Fe$_{x}$)$_{y}$As$_{2}$, $0\leq x\leq1$, $1.86\leq y \leq 2$, that elucidate changes to the magnetic order and lattice as $x$ is increased.  We find that A-type AFM order, with the ordered moments lying along the $c$ axis, persists for $x\lesssim0.12(1)$, and that both $T_{\textrm{N}}$ and the ordered moment are linearly suppressed with increasing $x$.  No dramatic changes to the lattice parameters are observed at $x\approx0.12$, and a smooth crossover from the cT to T phase occurs at higher $x$.  We compare our results with those for other compounds related to the $122$-Fe-pnictide superconductors, and discuss the relationship between changes in the lattice and magnetic order.

\section{EXPERIMENT}

Shiny plate-like single crystals of Ca(Co$_{1-x}$Fe$_{x}$)$_{y}$As$_{2}$ ($0\leq x\leq1$, $1.86\leq y \leq 2$) were synthesized by Sn-flux solution growth using Ca ($99.98\%$), Co ($99.998\%$), Fe ($99.998\%$) and As ($99.99999\%$) from Alfa Aesar.  Growth was initiated with stoichiometric Ca(Co$_{1-x}$Fe$_{x}$)$_{2}$As$_{2}$ ($x = 0$, $0.03$, $0.04$, $0.05$, $0.06$, $0.09$, $0.18$, $0.20$, $0.26$, $0.40$, $0.60$, $0.80$, $1$) and Sn in a $1:20$ sample to flux molar ratio.  Starting materials were placed in alumina crucibles and sealed in quartz tubes under $\approx 1/3$ atm of Ar.  After prereaction at $650$\,\degree C for $12$ hours, the materials were heated to $1050$\,\degree C at a rate of $40$\,\degree C$/$hour, held there for 20 hours, and then cooled to $700$\,\degree C at $-4$\,\degree C$/$hour.  Crystals were separated from the flux by decanting with a centrifuge at $700$\,\degree C.  Some single crystals were powdered for in-house x-ray diffraction measurements, and all of the samples were found to have the same body-centered tetragonal space group ($I4/mmm$) as the $x=0$ parent compound \cite{Quirinale_2013}.    

The magnetization $M$ was measured down to $T=1.8$~K and under applied magnetic fields of $H=0.1$ to $3$~T using a Quantum Design, Inc., Magnetic Property Measurement System in order to screen for a magnetic transition and determine any ordering temperatures.  Neutron diffraction experiments were subsequently carried out on selected samples using the HB-1A Fixed-Incident-Energy Triple-Axis Spectrometer at the High Flux Isotope Reactor, Oak Ridge National Laboratory. Measurements were made using a fixed incident neutron energy of $E=14.6$~meV, and collimators with divergences of $40^{\prime}$-$40^{\prime}$-$40^{\prime}$-$80^{\prime}$ were inserted before the pyrolytic graphite (PG) $(0~0~2)$ monochromator, between the monochromator and sample, between the sample and PG $(0~0~2)$ analyzer, and between the analyzer and detector, respectively.  Two PG filters were placed before the sample to suppress higher order harmonics present in the incident beam.  The samples were mounted with their  $(H~H~L)$ reciprocal-lattice planes coincident with the scattering plane, and were cooled down using either a He closed-cycle or an orange-type cryostat.  The masses of the samples range from $50$ to $100$ mg.  In this paper, momentum transfers are expressed in reciprocal-lattice units.

High-energy x-ray diffraction (HE-XRD) experiments were performed at end station $6$-ID-D at the Advanced Photon Source, Argonne National Laboratory, using an x-ray energy of $E=100.23$~keV.  Single-crystal samples were mounted on the cold finger of a He closed-cycle cryostat and cooled down to $T=6$~K, and He exchange gas was used to ensure thermal equilibrium.  The cryostat was mounted to the sample stage of a $6$-circle diffractometer, and either a MAR$345$ or a Pixirad-$1$ area detector was used to measure the diffracted x-rays transmitted through the sample.  The MAR$345$ image plate was positioned with its center aligned to the incident beam and was determined to be set back from the sample position by $2.832$~m  through measurement of a CeO$_{2}$ standard from the National Institute of Standards and Technology. In this configuration, the MAR$345$ recorded a diffraction pattern spanning a scattering angle of $|2\theta|\lesssim3.49\degree$.  The detector was operated with a pixel size of $100\times100~\mu\textrm{m}^{2}$, and diffraction patterns of the $(H~H~L)$ reciprocal-lattice plane were recorded using a $100\times100~\mu\textrm{m}^{2}$ incident beam while tilting the sample along two rocking angles.  Detailed studies of the temperature dependencies of the $(4~4~0)$ and $(0~0~8)$ Bragg reflections were recorded using the Pixirad-$1$.  For these measurements, the sample was rocked around an axis perpendicular to the incoming beam.   The detection element of the Pixirad-$1$ is comprised of a hexagonal array of pixels with a spacing of $60~\mu$m, and the incoming beam had a size of  $100\times100~\mu\textrm{m}^{2}$. 

\section{Results}
\subsection{Composition Analysis}
Previous x-ray diffraction, neutron diffraction, and wavelength-dispersive x-ray spectroscopy (WDS)  results for Sn-flux grown single-crystal samples of  CaCo$_{2}$As$_{2}$ show that $7(1)\%$ of the Co sites are vacant, resulting in a stoichiometry of CaCo$_{1.86(2)}$As$_{2}$ \cite{Quirinale_2013}.  To determine the chemical compositions of our Ca(Co$_{1-x}$Fe$_{x}$)$_{y}$As$_{2}$ samples, we performed either energy-dispersive x-ray spectroscopy (EDS) or WDS measurements.  Table~\ref{Tab1} lists the results, and from here on we will refer to samples by their values for $x$.

\begin{table}
 \caption{The chemical compositions of the Ca(Co$_{1-x}$Fe$_{x}$)$_{y}$As$_{2}$ samples as determined by EDS or WDS.  $x_{\textrm{nom}}$ is the nominal value for $x$ expected from the synthesis procedure.  \label{Tab1}}
\begin{ruledtabular}
 \begin{tabular}{@{\hspace{3em}} d{1.3}d{1.3} d{1.3}@{\hspace{3em}}  }
\multicolumn{1}{@{\hspace{3em}} c}{$x_{\textrm{nom}}$}&\multicolumn{1}{ c}{$x$}&\multicolumn{1}{c@{\hspace{3em}} }{$y$}\\
\colrule
0.00& 0.00& 1.86(2)\\
0.03& 0.043(2) & 1.88(4)\\
0.04& 0.057(2)& 1.88(3)\\
0.06& 0.068(2) & 1.89(6) \\
0.05& 0.091(8) & 1.90(4) \\
0.09& 0.104(4) & 1.90(4) \\
0.18& 0.19(4) & 1.94(5)\\
0.26& 0.25(2) & 1.96(1)\\
0.20& 0.35(2) & 2.00(6)\\
0.40& 0.48(2) & 2.00(4)\\
0.60& 0.67(2) & 2.0(1)\\
0.80& 0.89(2) & 2.0(1)\\
1.00& 1.00 & 2.00\\
 \end{tabular}
 \end{ruledtabular}
 \end{table}

\subsection{Magnetic Susceptibility}

\begin{figure}[]
	\centering
	\includegraphics[width=1.0\linewidth]{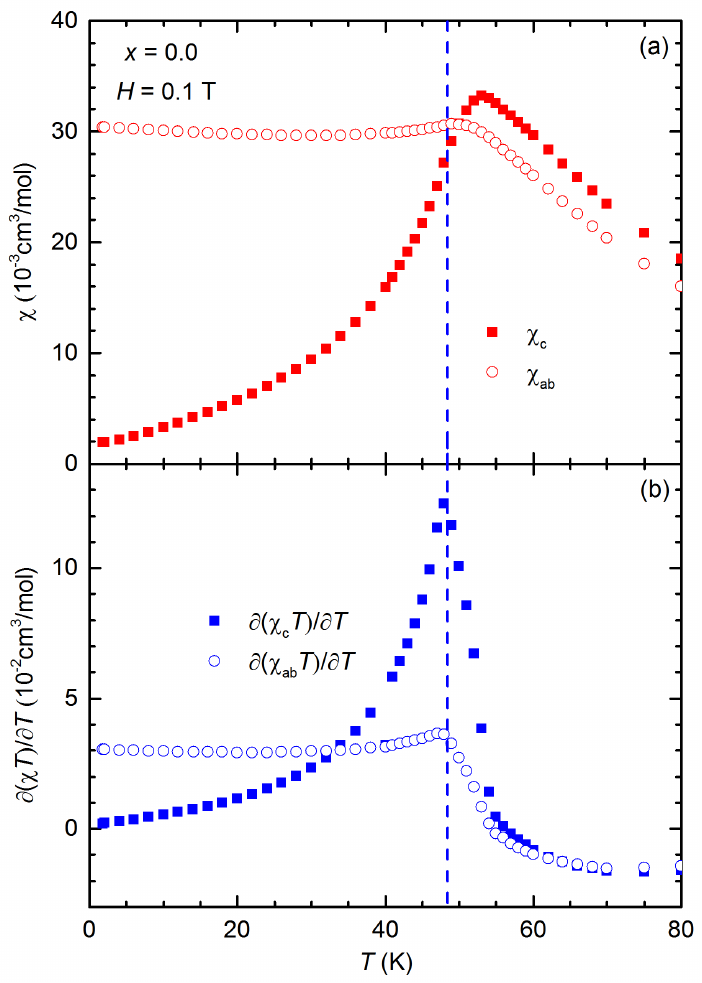}
	\caption{(Color online)  \label{Fig2} (a) Temperature dependence of the magnetic susceptibility of the $x=0$ sample for magnetic fields $H$ applied within the basal plane ($\chi_{\textrm{ab}}$) and along \textbf{c} ($\chi_{\textrm{c}}$).  These data also appear in Ref.~\onlinecite{Anand_2014}.  (b) The derivatives with respect to temperature of the products $\chi_{ab}T$ and $\chi_{\textrm{c}}T$.  The dashed line indicates the value of $T_{\textrm{N}}$ determined from the plots, as described in the text.}
\end{figure}

Figure~\ref{Fig2}(a) shows the magnetic susceptibility $\chi\equiv\frac{M}{H}$ versus temperature for $x=0$, from measurements made in the basal plane ($\chi_{\textrm{ab}}$) and along \textbf{c} ($\chi_{\textrm{c}}$).  Figure~\ref{Fig2}(b) shows plots of $\partial (\chi_{\textrm{ab}} T) / \partial T$ and $\partial (\chi_{\textrm{c}} T) / \partial T$.  From these plots, we use Fisher's method \cite{Fisher_1962} and determine that $T_{\textrm{N}}=48(2)$~K for $x=0$, which is slightly lower than the value of $T_{\textrm{N}}=52(1)$~K determined from previous neutron diffraction data \cite{Quirinale_2013}.  Note that $T_{\textrm{N}}$ may be readily determined using either  $\partial (\chi_{\textrm{ab}} T) / \partial T$ or $\partial (\chi_{\textrm{c}} T) / \partial T$.

\begin{figure}[]
	\centering
	\includegraphics[width=1.0\linewidth]{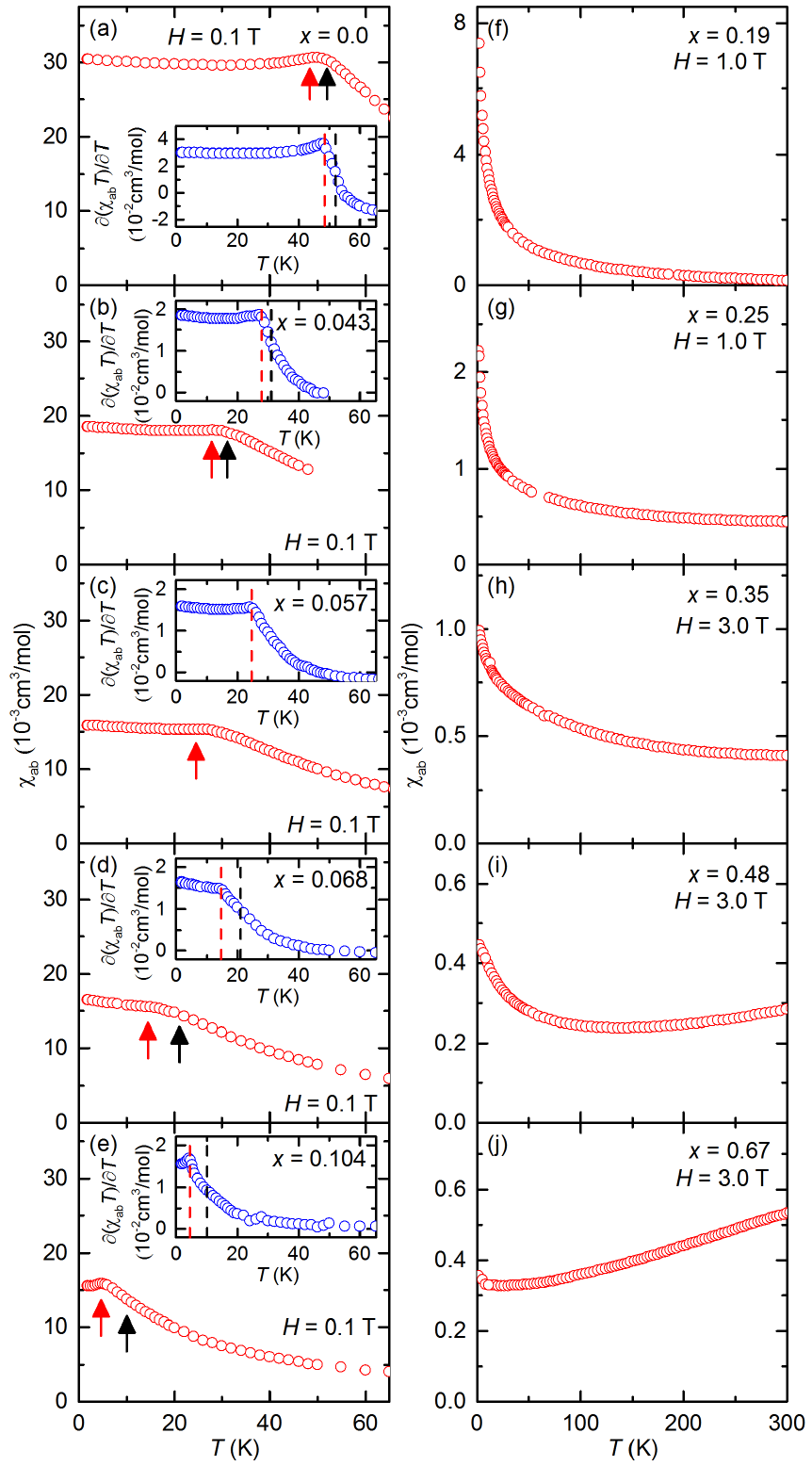}
	\caption{(Color online)  \label{Fig3} The magnetic susceptibilities versus temperature of the $x=0$ (a), $0.043$ (b), $0.057$ (c), $0.068$ (d), $0.104$ (e), $0.19$ (f), $0.25$ (g), $0.35$ (h), $0.48$ (i), and $0.67$ (j) samples for magnetic fields $H$ applied within the basal plane.  The insets show the derivatives with respect to temperature of $\chi_{ab}T$ for samples with an AFM transition.  Red arrows and lines indicate the values for $T_{\textrm{N}}$ determined from the data, whereas black arrows and lines indicate the values for $T_{\textrm{N}}$ determined by neutron diffraction.  The $\chi_{\textrm{ab}}(T)$ data for $x=0$  also appear in Ref.~\onlinecite{Anand_2014}.}
\end{figure}

Figure~\ref{Fig3} shows $\chi_{\textrm{ab}}(T)$ for the $x=0, 0.043$, $0.057$, $0.068$, $0.104$, $0.19$, $0.25$, $0.35$, $0.48$, and $0.67$ single-crystal samples, and the insets show corresponding plots of $\partial (\chi_{\textrm{ab}} T) / \partial T$ for samples which have an AFM transition.  The red arrows in Figs.~\ref{Fig3}(a)--\ref{Fig3}(e) denote the values of $T_{\textrm{N}}$ determined from $\partial (\chi_{\textrm{ab}} T) / \partial T$ using Fisher's method, whereas the black arrows indicate the the values determined from the neutron diffraction data presented below.  Figures~\ref{Fig3}(f)--\ref{Fig3}(j) reveal no features corresponding to a magnetic phase transition, but do show an upturn in $\chi(T)$ at low temperature. It is currently unclear if the upturn is due to magnetic impurities or intrinsic, and future investigations of the upturn are warranted \cite{Jia_2010, Jia_2011}.  Finally, none of the panels in Fig.~\ref{Fig3} show signs of diamagnetic behavior indicative of bulk superconductivity.

\subsection{Neutron Diffraction}
Neutron diffraction measurements were performed on single-crystal samples with $x=0$, $0.043$, $0.068$, $0.104$, $0.19$, and $0.25$ in order to determine the microscopic details of any magnetic order present at low temperature.  For a crystal possessing the body-centered-tetragonal space group $I4/mmm$, and oriented with its $(H~H~L)$ plane coincident with the scattering plane, the relevant reflection conditions for Bragg peaks due to the chemical lattice are $(H~H~L)$, $L$ even.   The A-type AFM order in the $x=0$ compound  breaks the body-centered symmetry, yielding additional Bragg peaks at $(H~H~L)$, $L$ odd, and is characterized by an AFM propagation vector $\bm{\tau}=(0~0~1)$ \cite{Quirinale_2013}.  The direction of the ordered moment is found by using the fact that neutron scattering is insensitive to a moment lying along the scattering vector \textbf{Q}.  Hence, since the ordered moment for the $x=0$ compound lies along the $c$ axis, magnetic Bragg peaks at $(0~0~L)$, $L$ odd, reciprocal-lattice positions are absent. 

\begin{figure}[]
	\centering
	\includegraphics[width=1.0\linewidth]{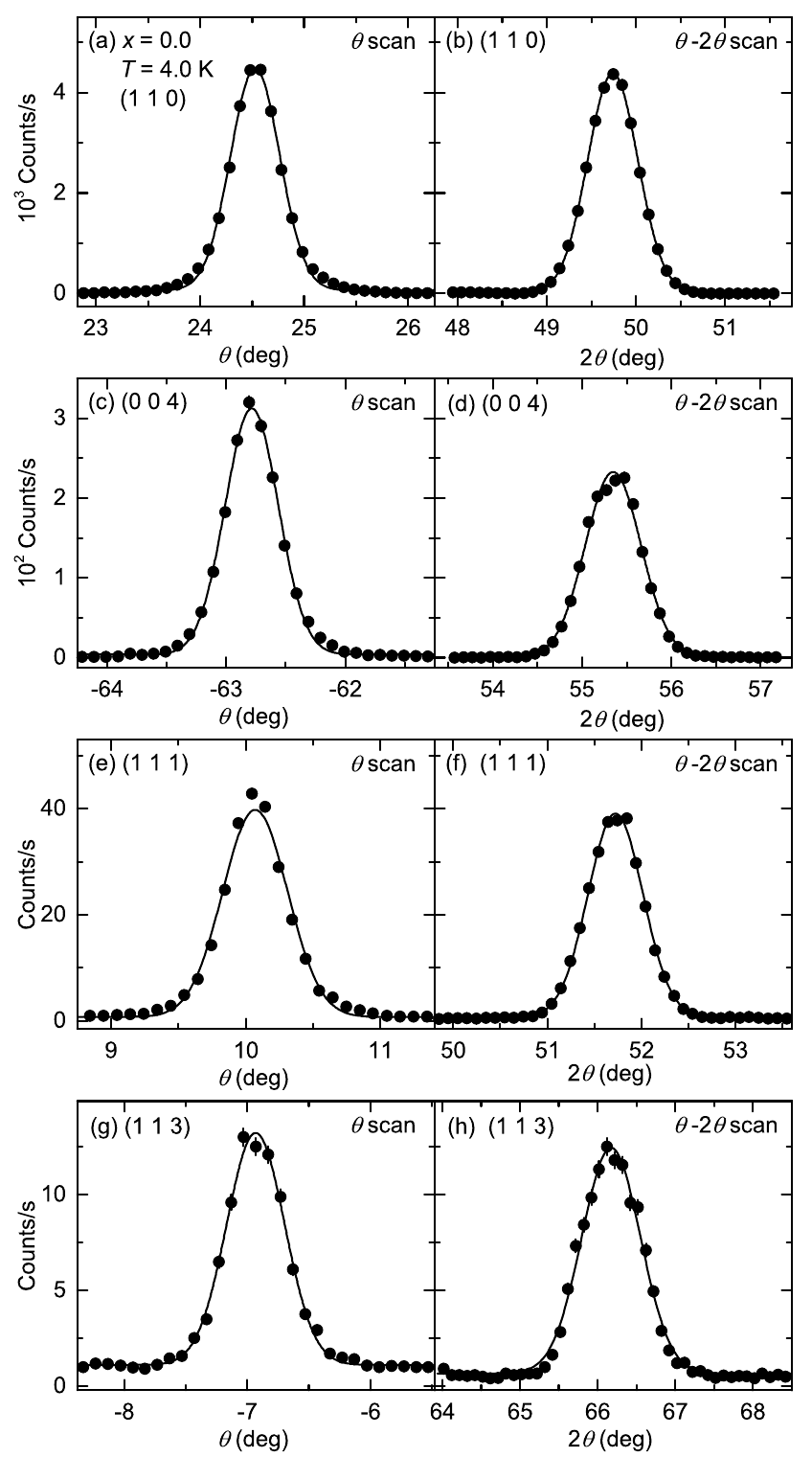}
	\caption{\label{Fig4} Neutron diffraction data for $x=0$ from $\theta$ and $\theta$-$2\theta$ scans through the $(1~1~0)$ [(a), (b)], $(0~0~4)$ [(c), (d)],  $(1~1~1)$ [(e), (f)], and $(1~1~3)$ [(g), (h)] Bragg peaks at $T=4$~K.  Solid lines are fits to Gaussian line shapes.}
\end{figure}

Figures~\ref{Fig4}(a)  and \ref{Fig4}(b) show data for the $x=0$ sample from $\theta$ and $\theta$-$2\theta$ scans (i.e. approximately transverse and longitudinal scans), respectively, through the $(1~1~0)$ Bragg peak at $T=4$~K.  Figures~\ref{Fig4}(c) and \ref{Fig4}(d) show similar data for the $(0~0~4)$ Bragg peak.  The smooth, sharp peaks in Figs.~\ref{Fig4}(a) and \ref{Fig4}(c)  have full widths at half maximum (FWHM), determined from fits to Gaussian line shapes, of $0.62(1)$\degree\ and $0.58(1)$\degree, respectively.  In Figs.~\ref{Fig4}(b) and \ref{Fig4}(d), the FWHM of the peaks are $0.671(5)$\degree\ and $0.716(5)$\degree, respectively, and the sharp peaks illustrate the quality of the single-crystal samples, since, presumably, the values of the lattice parameters should change with $x$.  The FWHM values listed above are typical for all of the samples used for the neutron diffraction measurements.

Figures~\ref{Fig4}(e) and \ref{Fig4}(f) show data from $\theta$ and $\theta$-$2\theta$ scans, respectively, through the $(1~1~1)$ magnetic Bragg peak, and Figs.~\ref{Fig4}(g) and \ref{Fig4}(h) show similar data for the $(1~1~3)$ magnetic Bragg peak.  Both of these peaks are due to the previously determined A-type AFM order \cite{Quirinale_2013}.  We determine a value for the ordered moment by comparing the integrated intensities of the $(1~1~1)$, $(1~1~3)$, and $(2~2~1)$ magnetic Bragg peaks and their calculated structure factors to the corresponding values for the $(0~0~2)$, $(1~1~0)$, $(0~0~4)$, $(1~1~2)$, $(1~1~4)$, and $(0~0~6)$ structural Bragg peaks.  We find that the ordered moment is $0.43(5)~\mu_{\textrm{B}}/$Co at $T=4$~K, which is consistent with the value of $\approx0.3~\mu_{\textrm{B}}/$Co from magnetization measurements \cite{Anand_2014} and the upper limit of $\approx0.6~\mu_{\textrm{B}}/$Co from previous neutron diffraction results \cite{Quirinale_2013}.  (The data for this calculation along with those for other $x$ are shown in Fig.~\ref{Fig12}.)

Next, we present results from experiments on an $x=0.068$ sample, and give in detail the procedures used to obtain the data and perform the analysis.  The methodology presented is the same one used for the rest of the samples measured by neutron diffraction.

\begin{figure}[]
	\centering
	\includegraphics[width=1.0\linewidth]{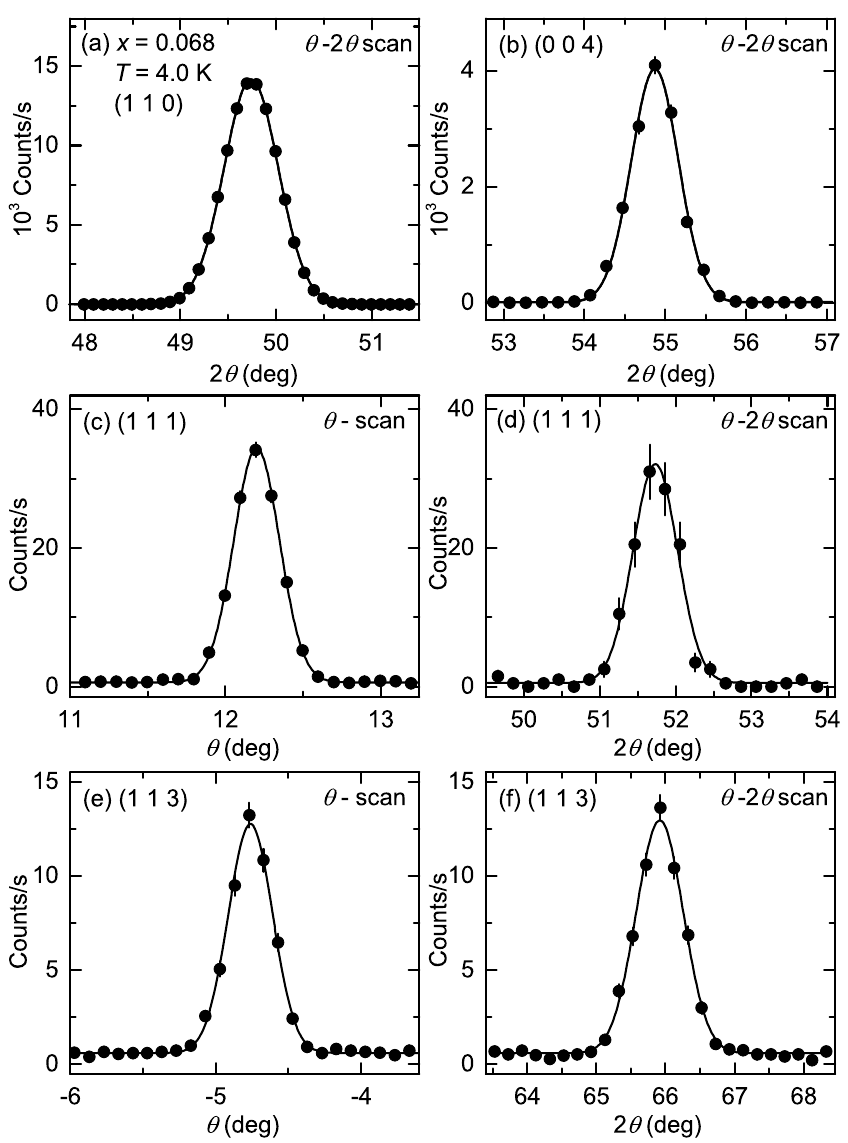}
	\caption{ \label{Fig5} Neutron diffraction data for $x=0.068$ from $\theta$-$2\theta$ scans through the $(1~1~0)$ (a) and $(0~0~4)$ (b) Bragg peaks at $T=4$~K, and $\theta$ and $\theta$-$2\theta$ scans through the $(1~1~1)$ [(c), (d)], and $(1~1~3)$ [(e), (f)] Bragg peaks at $T=4$~K.  Solid lines are fits to Gaussian line shapes.}
\end{figure}

Figures~\ref{Fig5}(a) and \ref{Fig5}(b) show data for $x=0.068$ from $\theta$-$2\theta$ scans through the $(1~1~0)$ and $(0~0~4)$ Bragg peaks at $T=4$~K.  The peaks have FWHM of $0.661(2)$\degree\ and $0.708(8)$\degree, respectively.  The widths are likely resolution limited, since the tightest collimator placed in the neutron beam has a divergence of $40^{\prime}$.   Figures~\ref{Fig5}(c) and \ref{Fig5}(d) show data from $\theta$ and $\theta$-$2\theta$ scans through the $(1~1~1)$ position and Figs.~\ref{Fig5}(e) and \ref{Fig5}(f) show data from similar scans through the $(1~1~3)$ position at $T=4$~K.  Analogous to the $x=0$ compound, Bragg peaks are present at the $(1~1~1)$ and $(1~1~3)$ reciprocal-lattice positions, which indicates that the body-centered symmetry present at room temperature is broken at low temperature. The FWHM of the peaks in Figs.~\ref{Fig5}(d) and \ref{Fig5}(f) are $0.66(3)$\degree\ and $0.81(2)$\degree, respectively, which are comparable to the values found for the $(1~1~0)$ and $(0~0~4)$ Bragg peaks, and  indicate that the peaks are due to long-range order.

\begin{figure}[]
	\centering
	\includegraphics[width=1.0\linewidth]{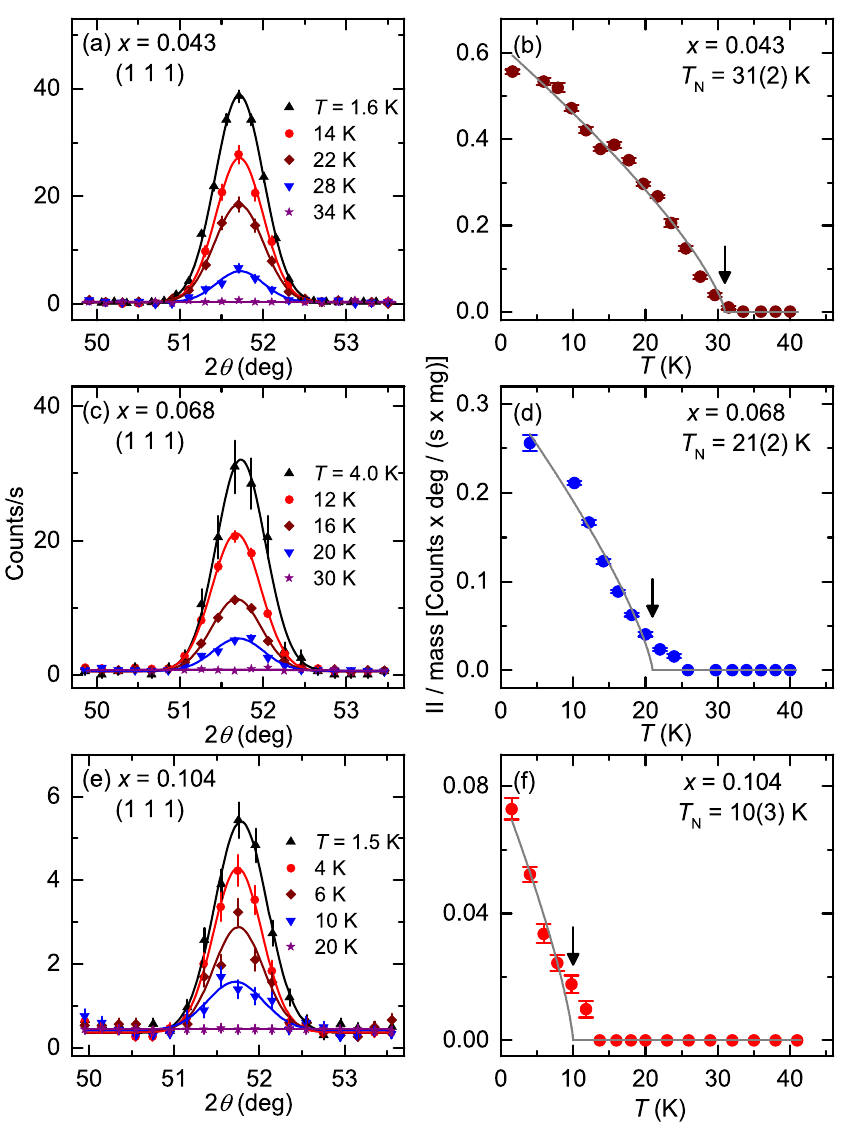}
	\caption{(Color online)  \label{Fig6} Neutron diffraction data from $\theta$ - 2$\theta$ scans through the $(1~1~1)$ Bragg peak and the temperature dependence of the integrated intensity (II) of the peak for $x=0.043$ [(a), (b)] , $x = 0.068$  [(c), (d)], and $ x = 0.104$ [(e), (f)].  Lines in (a), (c), and (e) are fits to Gaussian line shapes, and lines in (b), (d), and (f) are guides to the eye.  Arrows indicate the values for $T_{\textrm{N}}$ determined from the data.}
\end{figure}

Data for $x=0.068$ from $\theta$-$2\theta$ scans through the $(1~1~1)$ Bragg peak at different temperatures are plotted in Fig.~\ref{Fig6}(c), and Fig.~\ref{Fig6}(d) shows the temperature evolution of the peak's integrated intensity, which is determined by fitting the peaks in Fig.~\ref{Fig6}(c) to Gaussian line shapes. The values for the integrated intensity have been normalized by the sample mass.  Upon cooling, the $(1~1~1)$  peak appears below $T\approx 25$~K, which is similar to the temperature at which $\chi(T)$ deviates from its high-temperature behavior [see Fig.~\ref{Fig3}(d)].  Hence, we associate the appearance of the peak with the development of magnetic order.  The integrated intensity versus temperature curve has a high-temperature tail, which is likely due to short-range magnetic correlations associated with the phase transition, but compositional disorder within the samples may also contribute.  Similar shaped curves are observed for other compounds related to the Fe-pnictide superconductors, such as Ba(Fe$_{1-x}$Mn$_{x}$)$_{2}$As$_{2}$ \cite{Kim_2010}.  By extrapolating the expected behavior of the magnetic order parameter from low temperature, we determine that $T_{\textrm{N}}=21(2)$~K.  Figures~\ref{Fig6}(a), \ref{Fig6}(b), \ref{Fig6}(e), and \ref{Fig6}(f) show similar data for $x=0.043$ and $0.104$, which will be discussed later.

Whereas measurements of many Bragg peaks are necessary to uniquely determine a magnetic structure, the presence of both the $(1~1~1)$ and $(1~1~3)$  Bragg peaks for $x=0.068$ suggests that the magnetic structure is similar to the A-type AFM order in $x=0$.  To test this, we made various diffraction measurements: ($1$) to determine if other magnetic Bragg peaks consistent with A-type AFM order exist, ($2$) to determine if the ordered moment lies solely along the $c$ axis, and ($3$) to search for scattering consistent with the development of the stripe-type AFM order in CaFe$_{2}$As$_{2}$ \cite{Goldman_2009,Kreyssig_2008} and other compounds related to the Fe-pnictide superconductors \cite{Johnston_2010, Canfield_2010,Lynn_2009, Paglione_2010}.  Total-energy calculations for SrCo$_{2}$As$_{2}$ also indicate that a FM ground state lies in close proximity to stripe-type AFM, A-type AFM, and nonmagnetic ground states \cite{Jayasekara_2015}.  Therefore, we also looked for evidence indicating that FM order develops with increasing $x$.  Regarding point ($1$), a Bragg peak also occurs at the $(2~2~1)$ position at $T=4$~K, which is another position consistent with the A-type order in $x=0$.  The peak is absent at $T=50$~K.

\begin{figure}[]
	\centering
	\includegraphics[width=1.0\linewidth]{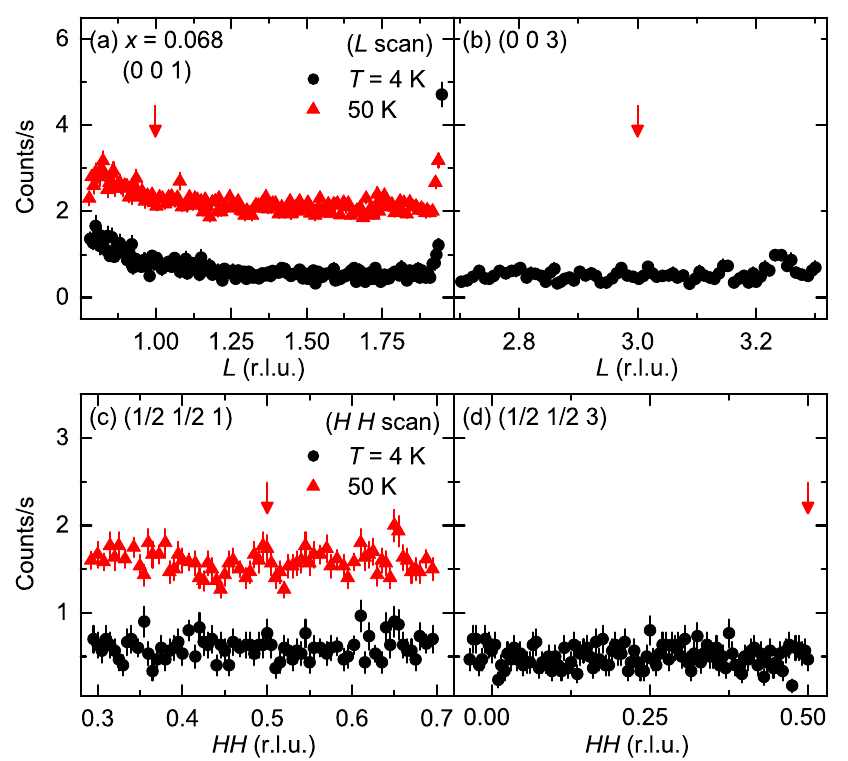}
	\caption{(Color online)  \label{Fig7} Neutron diffraction data for $x=0.068$  from \textbf{Q} scans through the $(0~0~1)$ (a), $(0~0~3)$ (b), $(\frac{1}{2}~\frac{1}{2}~1)$ (c), and $(\frac{1}{2}~\frac{1}{2}~3)$ (d) reciprocal-lattice positions.  In (a) and (c), data taken at $T=50$~K are shown in red and offset by $1.5$ and $1$ counts/s, respectively.  Arrows in (a) and (b) indicate the positions for peaks that would correspond to A-type AFM order with a component of the ordered moment lying in the $ab$ plane.  Arrows in (c) and (d) indicate the positions for peaks that would correspond to the stripe-type AFM order possessed by CaFe$_{2}$As$_{2}$.  Data in (a), (b), and (d) are for a beam monitor value corresponding to a counting time of $30$~s per point, and data in (c) are for a monitor value corresponding to $60$~s per point.}
\end{figure}

In regards to point ($2$), data for $x=0.068$ from longitudinal scans through the $(0~0~1)$ and $(0~0~3)$  positions are shown in Figs.~\ref{Fig7}(a) and \ref{Fig7}(b), respectively.  Whereas the tail of the $(0~0~2)$ peak is visible  in Fig.~\ref{Fig7}(a) at both $T=50$ and $4$~K, there is no evidence of a Bragg peak at $(0~0~1)$.  Similarly, no Bragg peak is seen in Fig.~\ref{Fig7}(b) at $T=4$~K for the $(0~0~3)$ position.  The existence of these peaks would be consistent with a component of the ordered magnetic moment lying in the $ab$ plane.  Hence, since the peaks are absent, the ordered moment lies along the $c$ axis.

Addressing point (3), Figs.~\ref{Fig7}(c) and \ref{Fig7}(d) show data for $x=0.068$ from scans through the $(\frac{1}{2}~\frac{1}{2}~1)$ and $(\frac{1}{2}~\frac{1}{2}~3)$ reciprocal-lattice positions, respectively, which correspond to $\bm{\tau}_{\textrm{\textbf{st}}}$.  Magnetic Bragg peaks do not occur at either position, which indicates that stripe-type AFM order similar to that in CaFe$_{2}$As$_{2}$ does not occur at $T=4$~K. In addition, the absence of any magnetic Bragg peaks in Fig.~\ref{Fig7} rules out the existence of incommensurate long-range magnetic order with a propagation vector consistent with the investigated values of \textbf{Q}.  

\begin{figure}[]
	\centering
	\includegraphics[width=1.0\linewidth]{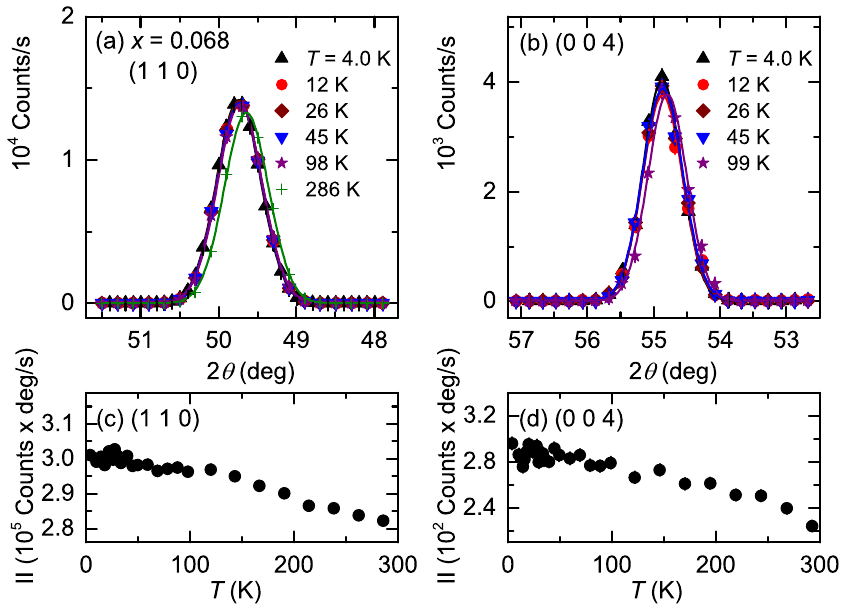}
	\caption{(Color online)  \label{Fig8} Neutron diffraction data for $x=0.068$ from $\theta$ - 2$\theta$ scans through the $(1~1~0)$ (a) and $(0~0~4)$ (b) Bragg peaks at various temperatures. Lines are fits to Gaussian line shapes.  The integrated intensities (II) determined from the fits are plotted versus temperature in (c) [$(1~1~0)$] and (d) [$(0~0~4)$].}
\end{figure}

Finally, for $x=0.068$, we examine the temperature dependence of certain Bragg peaks with indices $(H~H~L)$, $L$ even, in order to look for evidence of FM order.  Figure~\ref{Fig8} shows data from $\theta$-$2\theta$ scans through the $(1~1~0)$ [Fig.~\ref{Fig8}(a)] and $(0~0~4)$ [Fig.~\ref{Fig8}(b)] Bragg peaks at various temperatures.  Due to the reciprocal-lattice positions of the peaks, and the fact that neutron diffraction is sensitive to a magnetic moment's component perpendicular to \textbf{Q}, the measurements cover the possible development of an ordered FM moment with components in either the $ab$ plane or along the $c$ axis.  The temperature evolution of the integrated intensities of the $(1~1~0)$ and $(0~0~4)$ Bragg peaks are shown in Figs.~\ref{Fig8}(c) and \ref{Fig8}(d), respectively.  Both datasets vary smoothly with $T$, and slightly increase with decreasing temperature, likely due to the temperature dependence of their Debye-Waller factors.  Nevertheless, there are no sharp changes to the shapes nor in the  integrated intensities of both Bragg peaks as $T$ is lowered, and we find no significant indication for the development of FM order down to $T=4$~K.

\begin{figure}[]
	\centering
	\includegraphics[width=1.0\linewidth]{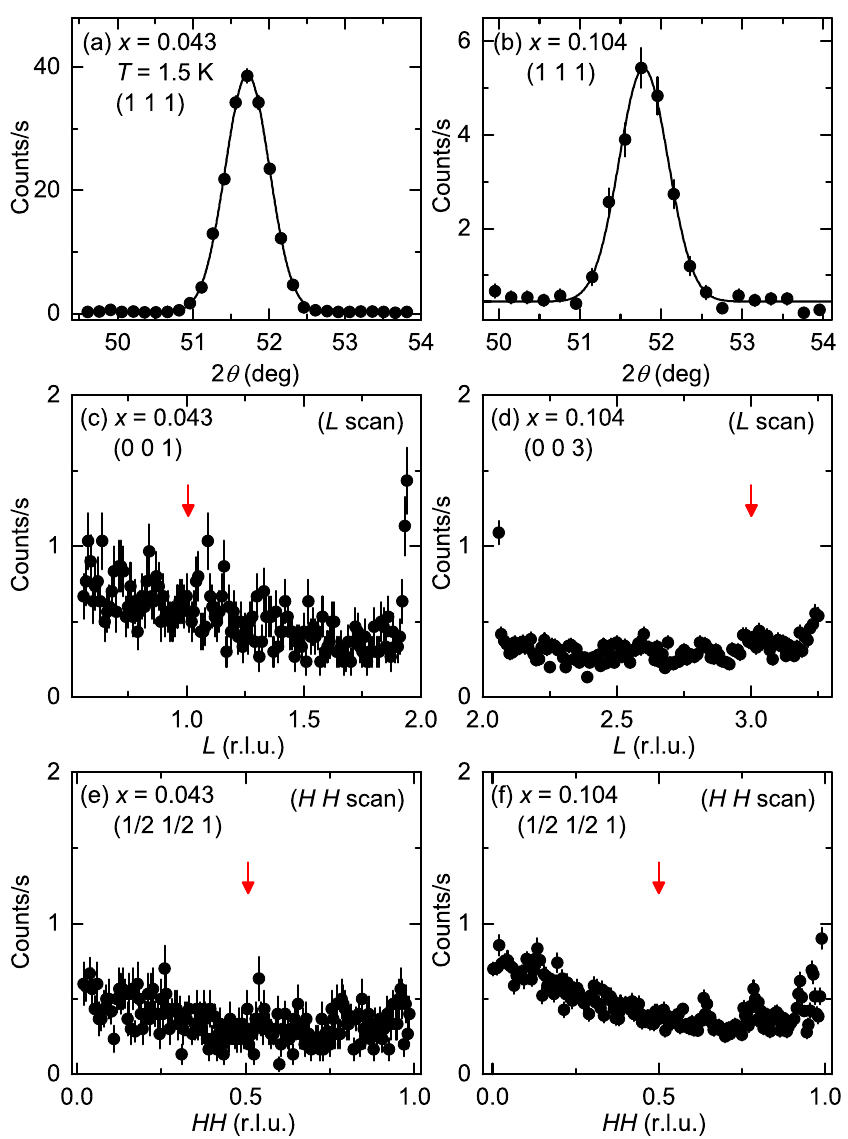}
	\caption{\label{Fig9} Neutron diffraction data from scans across possible magnetic Bragg peak positions at $T=1.5$~K for $x=0.043$ and $0.104$.  (a), (b) Data from $\theta$-$2\theta$ scans of the $(1~1~1)$ magnetic Bragg peak for $x=0.043$ (a)  and $x=0.104$ (b).  Lines are fits to Gaussian line shapes.   (c), (d)  Data from longitudinal scans through the $(0~0~1)$ reciprocal-lattice position for $x=0.043$ (c), and through the $(0~0~3)$ position for $x=0.104$ (d).  (e), (f) Data from scans through the $(\frac{1}{2}~\frac{1}{2}~1)$ position for $x=0.043$ (e) and $0.104$ (f).  Arrows indicate either the expected positions for peaks arising due to a component of the ordered moment lying in the $ab$ plane [(c), (d)] or for peaks corresponding to the stripe-type AFM order possessed by CaFe$_{2}$As$_{2}$ [(e), (f)].   Data in (c) and (e) are for a beam monitor value corresponding to a counting time of $30$~s per point, and data in (d) and (f) are for a monitor value corresponding to $180$~s per point.  The small peaks in (f) are ruled out as being magnetic Bragg peaks by results from complimentary measurements made above $T_{\textrm{N}}$, or are ruled out as being due to the single-crystal sample by results from $\theta$ scans across their positions.}
\end{figure}

\begin{figure}[]
	\centering
	\includegraphics[width=1.0\linewidth]{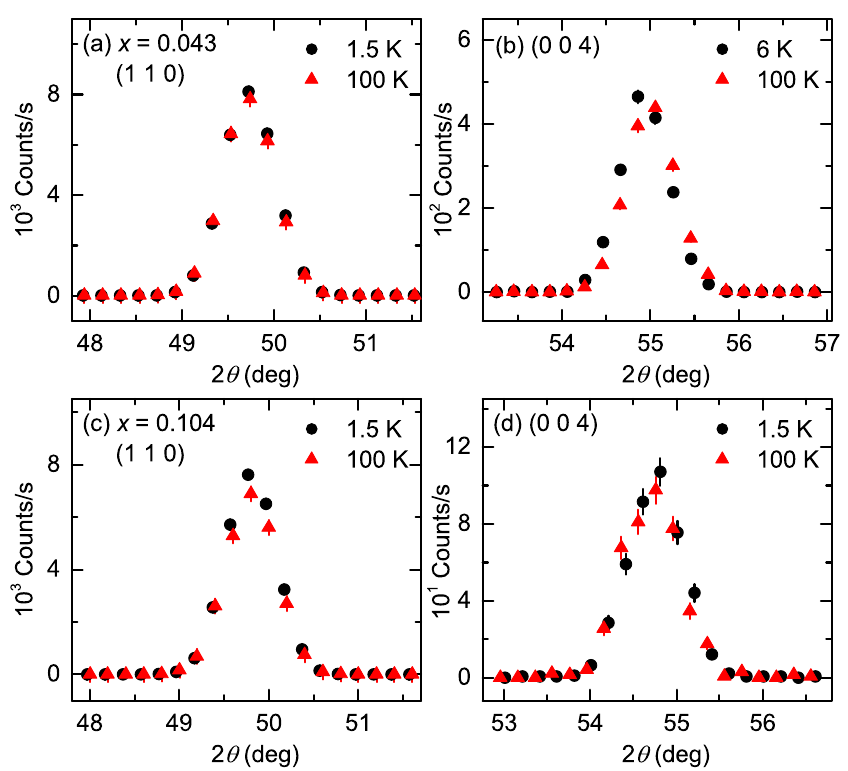}
	\caption{(Color online)  \label{Fig10} Neutron diffraction data from $\theta$-$2\theta$ scans through the $(1~1~0)$ [(a) and (c)] and $(0~0~4)$ [(b) and (d)] Bragg peaks for $x=0.043$ and $0.104$ at the temperatures indicated.  Data for $x=0.043$ are shown in (a) and (b), and data for $x=0.104$ are shown in (c) and (d).}
\end{figure}

Similar data to those presented in Figs.~\ref{Fig4}--\ref{Fig8} were recorded for the $x=0.043$ and $0.104$ samples, and the salient data are shown in Figs.~\ref{Fig9} and \ref{Fig10}.   Figures~\ref{Fig9}(a) and \ref{Fig9}(b) show data from $\theta $-$2\theta$ scans through the $(1~1~1)$ reciprocal-lattice positions for $x=0.043$ and $0.104$, respectively, and illustrate that peaks are found for both samples which are consistent with A-type AFM order.  The FWHM of the peaks are $0.671(7)$\degree\ and $0.72(4)$\degree\ for $x=0.043$ and $0.104$, respectively, which indicate that the peaks correspond to long-range magnetic order. Figure~\ref{Fig9}(c) shows data from a longitudinal scan through the $(0~0~1)$ position for the $x=0.043$ sample and Fig.~\ref{Fig9}(d) shows data through the $(0~0~3)$ position for $x=0.104$.  The absence of Bragg peaks in these data means that the ordered moment lies along the $c$ axis for both values of $x$.  Finally, Figs.~\ref{Fig9}(e) and \ref{Fig9}(f) show data from scans through the $(\frac{1}{2}~\frac{1}{2}~1)$ positions for $x=0.043$ and $0.104$, respectively.  No peak is seen for either value of $x$, which illustrates that the stripe-type AFM order present in CaFe$_{2}$As$_{2}$ does not occur in these samples.  Similarly, Fig.~\ref{Fig10} shows that the $(1~1~0)$ and $(0~0~4)$ Bragg peaks do not change between $T=100$~K and base temperature, which indicates that there is also no significant evidence for the development of FM order in these samples.

The temperature evolution of the $(1~1~1)$ magnetic Bragg peaks and the magnetic order parameters for the $x=0.043$ and $0.104$ samples are shown in Figs.~\ref{Fig6}(a), \ref{Fig6}(b), \ref{Fig6}(e), and \ref{Fig6}(f).  Using the same procedure described above to determine $T_{\textrm{N}}$ for $x=0.068$, we find that  $T_{\textrm{N}}=31(2)$~K for $x=0.043$ and $T_{\textrm{N}}=10(3)$~K for $x=0.104$.  These values are similar to those indicated by the red arrows in Figs.~\ref{Fig3}(b) and \ref{Fig3}(e), and we conclude that $T_{\textrm{N}}$ decreases with increasing $x$.  In addition, since the ordinates of Figs.~\ref{Fig6}(b), \ref{Fig6}(d), and \ref{Fig6}(f) give the integrated intensity of the $(1~1~1)$ peak normalized by the sample mass, and since the integrated intensity is proportional to the square of the ordered magnetic moment, the decrease in the base temperature values of the normalized integrated intensity with increasing $x$ indicates that the ordered moment is suppressed as Fe is substituted for Co.  We will return to these points with a subsequent figure.

\begin{figure}[]
	\centering
	\includegraphics[width=1.0\linewidth]{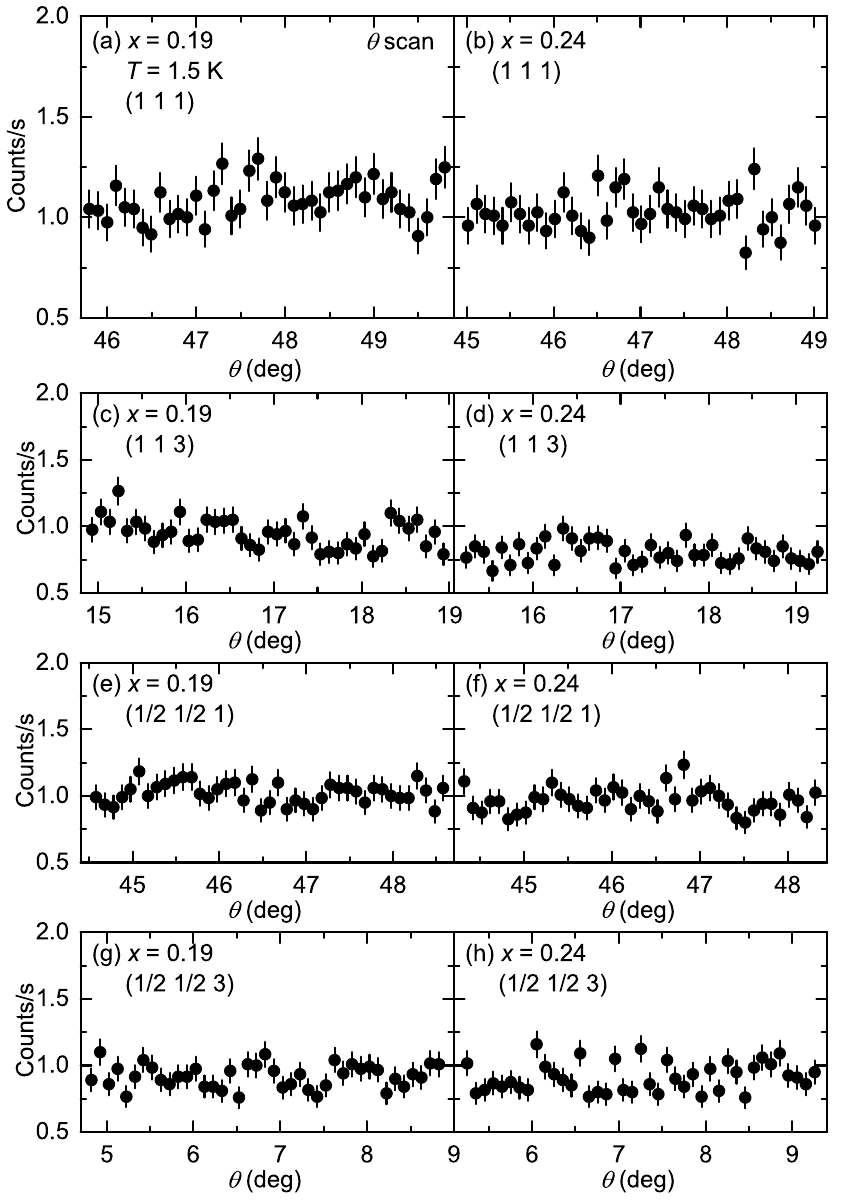}
	\caption{\label{Fig11} Neutron diffraction data from $\theta$ scans through the  $(1~1~1)$ [(a), (b)] , $(1~1~3)$  [(c), (d)],  $(\frac{1}{2}~\frac{1}{2}~1)$ [(e), (f)], and $(\frac{1}{2}~\frac{1}{2}~3)$ [(g), (h)]  reciprocal-lattice positions for $x = 0.19$ [(a), (c), (e), (g)] and $x=0.25$ [(b), (d), (f), (h)] at $T = 1.5$~K.   All of the data shown are for a beam monitor value corresponding to a counting time of $120$~s per point.}
\end{figure}

\begin{figure}[]
	\centering
	\includegraphics[width=1.0\linewidth]{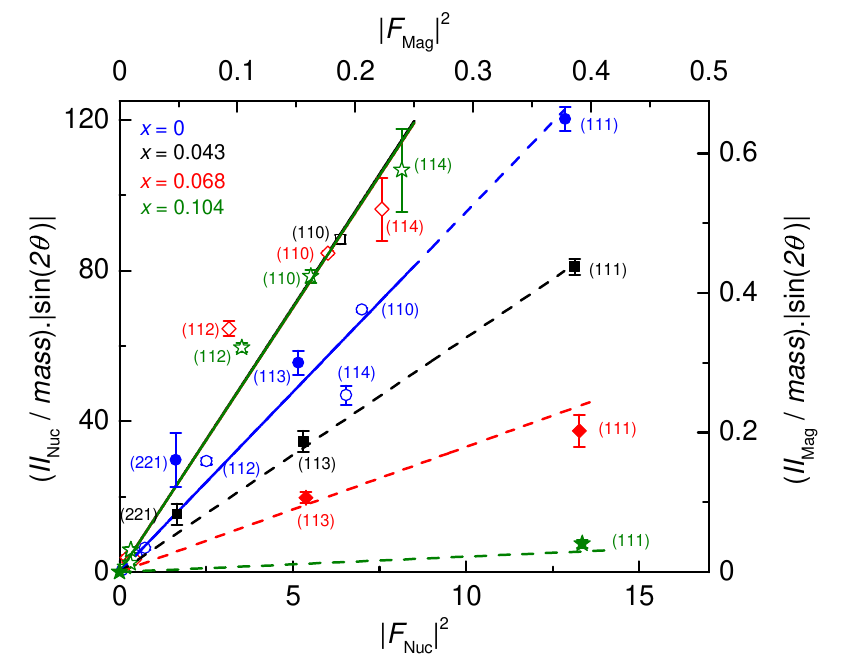}
	\caption{(Color online)  \label{Fig12} Integrated intensities of the measured structural (left axis, open symbols) and magnetic (right axis, closed symbols) Bragg peaks for the $x=0$, $0.043$, $0.068$, and $0.104$ samples plotted versus the square of their calculated structure factors.  The bottom axis corresponds to the square of the chemical structure factor, whereas the top axis corresponds to the square of the magnetic structure factor calculated for A-type AFM order with an ordered moment of $\mu=1~\mu_{\textrm{B}}$ laying along \textbf{c}.  The integrated intensities have been normalized to the sample mass and corrected by the Lorentz factor.  Solid (dashed) lines show linear fits to the structural (magnetic) data.}
\end{figure}

Figure~\ref{Fig11} shows data for $x=0.19$ and $0.25$ from $\theta$ scans performed at $T=1.5$~K through reciprocal-lattice positions corresponding either to A-type or stripe-type order.  Data for the $(1~1~1)$ position are shown in Fig.~\ref{Fig11}(a) for $x=0.19$, and  in Fig.~\ref{Fig11}(b) for $x=0.25$. No magnetic Bragg peaks occur in these data.  Similarly, Figs.~\ref{Fig11}(c) and \ref{Fig11}(d) show that the $(1~1~3)$ peak is absent for both values of $x$.  This indicates that the A-type AFM order found for lower values of $x$ is absent, at least within the sensitivity of our measurement.  Next, Figs.~\ref{Fig11}(e) and \ref{Fig11}(f) show data for the $(\frac{1}{2}~\frac{1}{2}~1)$ position for $x=0.19$ and $0.25$, respectively, and Figs.~\ref{Fig11}(g) and \ref{Fig11}(h) show data for the $(\frac{1}{2}~\frac{1}{2}~3)$ position for $x=0.19$ and $0.25$, respectively.  No Bragg peak is found in any of these figures, which indicates that the stripe-type AFM order present for $x=1$ is absent for $x=0.19$ and $0.25$.  We can roughly estimate the minimum ordered moment detectable by our measurements by using the data for $x=0$, from which we find that an ordered moment of $\mu=0.43(5)~\mu_{\textrm{B}}/$Co corresponds to a height of the $(1~1~1)$ magnetic Bragg peak of $\approx40$~counts/s.  Thus,  a peak with a height of $0.2$~counts$/$s, which is an estimate for the ability to distinguish a peak in the data in Fig.~\ref{Fig11}, would correspond to an ordered moment of $\approx0.03~\mu_{\textrm{B}}/$Co.

The integrated intensities of the measured structural and magnetic Bragg peaks normalized by the appropriate sample mass and Lorentz factor are plotted versus the squares of their respective structure factors for $x=0$, $0.043$, $0.068$, and $0.104$ in Fig.~\ref{Fig12}.  For the magnetic structure factor calculation, it has been assumed that A-type AFM order occurs with moments of $\mu=1~\mu_{\textrm{B}}$ lying along the $c$ axis.   The solid and dashed lines show linear fits to the structural and magnetic data, respectively.  For a given sample, the ratio of the slope of the line for the magnetic data to the slope of the line for the structural data equals the square of the ordered moment.  Thus, the decrease in slope with increasing $x$ for the magnetic data illustrates the decrease in $\mu$ with increasing $x$.  Note that there are several points at $0$ for the magnetic data which correspond to the magnetic structure factors for the $(0~0~1)$ and $(0~0~3)$ positions.  Despite the few number of magnetic Bragg peaks measured for each sample, the fits show that the data are consistent with A-type AFM order with the ordered moments lying along \textbf{c}.  The rise in slope of the fits to the structural data between $x=0$ and $0.043$  simply reflects the fact that the neutron scattering length for Fe is larger than the scattering length for Co.

\begin{figure}[]
	\centering
	\includegraphics[width=1.0\linewidth]{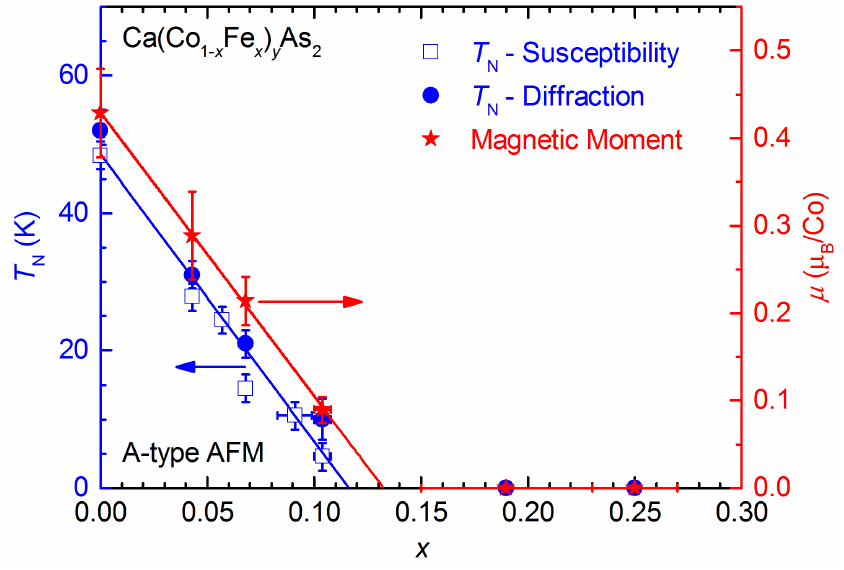}
	\caption{(Color online)  \label{Fig13} Magnetic phase diagram and ordered magnetic moment $\mu$ for low values of $x$.  Lines are fits to the data as described in the text.}
\end{figure}

Figure~\ref{Fig13} summarizes the results from the neutron diffraction and magnetic susceptibility experiments.   The left ordinate gives $T_{\textrm{N}}$, and the right ordinate gives the value for $\mu$.  Both $\mu$ and $T_{\textrm{N}}$ decrease with increasing $x$ in similar fashions.  A linear fit to the non-zero values for $T_{\textrm{N}}(x)$ determined from both the neutron diffraction and susceptibility data gives a slope of  $\frac{dT_{\textrm{N}}}{dx}=-418(27)$~K and an $x$ intercept of $x=0.116(8)$.  A similar fit to $\mu(x)$ yields $\frac{d\mu}{dx}=-3.24(5)\mu_{\textrm{B}}/$Co and an $x$ intercept of $x=0.132(2)$.  The fits are shown in Fig.~\ref{Fig13} as blue and red lines, respectively.  From the mean of the horizontal intercepts, we find that the A-type order is completely suppressed at $x=0.12(1)$.

\subsection{X-ray Diffraction}
\begin{figure}[]
	\centering
	\includegraphics[width=1.0\linewidth]{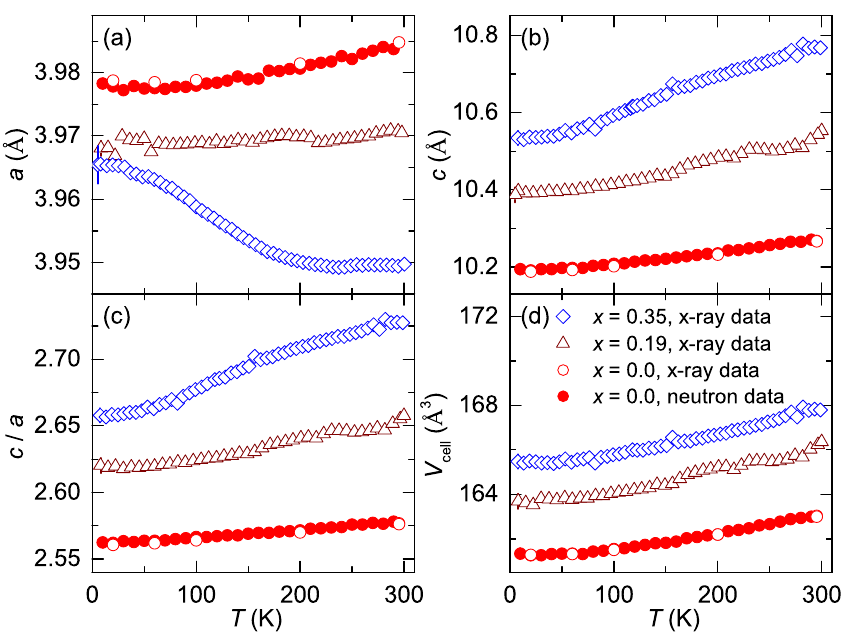}
	\caption{(Color online)  \label{Fig14} Temperature evolution of the $a$ (a) and $c$ (b) lattice parameters, the ratio $c/a$ (c), and $V_{\textrm{cell}}$ (d) for $x=0$, $0.18$, and $0.35$.  Data shown are from either neutron or high-energy x-ray diffraction experiments.  The uncertainty in the data is either within the symbol size or indicated by a representative error bar.  Data for $x=0$ are reproduced from Ref.~\onlinecite{Quirinale_2013}.}
\end{figure}

Figure~\ref{Fig14} shows the temperature dependence of the lattice parameters for samples with $x=0$, $0.19$, and $0.35$.  In Fig.~\ref{Fig14}(a),  $a$ increases between $T=10$ and $290$~K by only $0.1\%$ for $x=0$, remains virtually unchanged for $x=0.19$, and decreases by $0.4\%$ for $x=0.35$.  On the other hand, Fig.~\ref{Fig14}(b) shows that $c$ increases with increasing $T$ for all three samples: $0.7\%$ for $x=0$, $1.6\%$ for $x=0.19$, and $2.2\%$ for $x=0.35$.  The concomitant changes of $c/a$ and $V_{\textrm{cell}}$ with increasing temperature are shown in Figs.~\ref{Fig14}(c) and \ref{Fig14}(d), respectively.  Both quantities increase with increasing $T$ due to the much larger change in $c$ than $a$.

\begin{figure}[]
	\centering
	\includegraphics[width=1.0\linewidth]{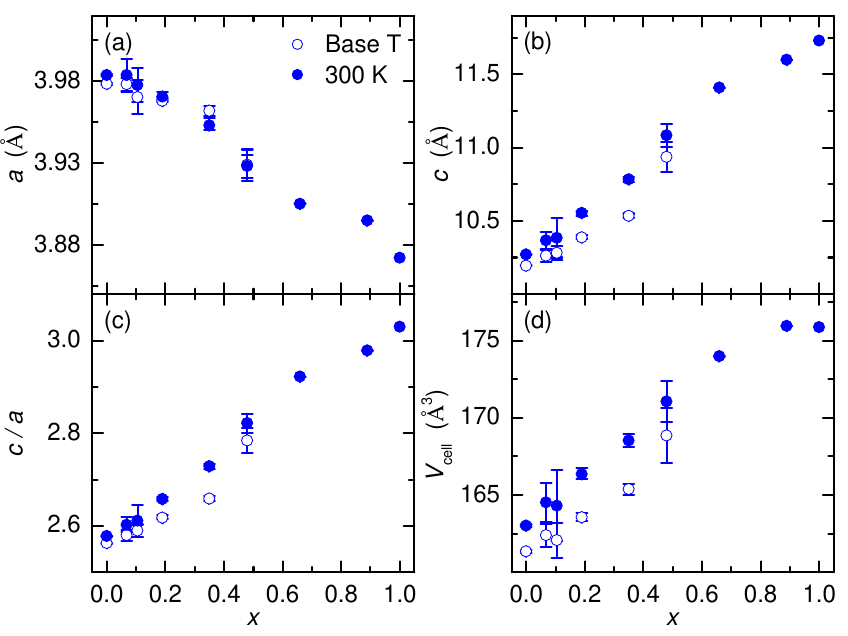}
	\caption{(Color online)  \label{Fig15} Evolution of the $a$ (a) and $c$ (b) lattice parameters, the ratio $c/a$ (c), and $V_{\textrm{cell}}$ (d) with Fe doping concentration $x$. Filled symbols denote data at $T=300$~K and open symbols denote data at base temperature ($4\leq T \leq 10$~K).  Data are shown from neutron diffraction, high-energy x-ray diffraction, and laboratory-based x-ray diffraction experiments.}
\end{figure}

Figure~\ref{Fig15} shows $a$, $c$, $c/a$, and $V_{\textrm{cell}}$ at $T=300$~K and base temperature ($4\leq T \leq 10$~K) for $0\leq x \leq 0.48$, and at $T=300$~K for $0.67\leq x \leq 1$.  For each temperature, $a$ monotonically decreases with increasing $x$ whereas $c$, $c/a$, and $V_{\textrm{cell}}$ smoothly increase.  The limited data taken at base temperature may show a sharp change in $c/a$ between $x=0.35$ and $0.48$, but a firm conclusion cannot be drawn from the data.  Nevertheless, there is no obvious feature at base-temperature which would corresponds to the disappearance of A-type AFM order at $x=0.12(1)$, and a smooth crossover between the cT and T phases occurs with increasing $x$.  This is somewhat surprising considering that a pressure-induced first-order transition between the T and cT phases occurs in CaFe$_{2}$As$_{2}$ \cite{Kreyssig_2008,Goldman_2008}.  On the other hand, substituting Fe for Co changes not only the chemical structure of the compound, but also the electronic band structure.  We further discuss this point below.

\section{Discussion}
The substitution of Fe for Co in CaCo$_{1.86}$As$_{2}$ dopes holes in to the compound, whereas the substitution of Co for Fe in CaFe$_{2}$As$_{2}$ dopes electrons.  In a rigid-band approximation, such doping shifts the Fermi energy $E_{\textrm{F}}$ and changes the partial density of states (DOS) at the Fermi level of the $3d$ bands of the transition metals, potentially affecting the magnetism.  For BaFe$_{2}$As$_{2}$ this scheme has been ruled out, because data for Ba(Fe$_{1-x}M_{x}$)$_{2}$As$_{2}$, in which $M=$~Co, Ni, Cu, or Co/Cu mixtures were systematically substituted for Fe, show that the positions of both the orthorhombic and AFM phase lines scale with $x$, not the number of electrons added via doping \cite{Canfield_2010, Canfield_2009}.  On the other hand, for $x$ up to at least $x=0.114$, $c/a$ changes with the number of extra electrons added per transition metal site in a similar manner for each type of transition metal used. For Ba(Fe$_{1-x}$Co$_{x}$)$_{2}$As$_{2}$, ARPES data also rule out a strict rigid-band approximation scenario, since the Fermi-surface hole pockets present for values of $x$ corresponding to AFM order vanish at a Lifshitz transition tied to the value of $x$ for which superconductivity first occurs \cite{Liu_2010}.

The changes between the T and cT phases for compounds such as CaFe$_{2}$As$_{2}$, with general formula $AT_{2}X_{2}$ and the ThCr$_{2}$Si$_{2}$ structure, may be described  in terms of the valence and  spacing of the $T_{2}X_{2}$ and $A$ layers \cite{Hoffman_1985}.  In such a description, compounds in the T phase have layers with valence assignments of $A^{+2}$ and ($T_{2}X_{2}$)$^{-2}$, and the distance separating two $T_{2}X_{2}$ layers in a unit cell is great enough that essentially no bonding occurs between them.  In the cT phase, the interlayer spacing decreases to an amount comparable to the distance necessary for a covalent bond to form between two $X$ anions.  This results in interlayer bonds developing along $c$ between $X$ anions in adjacent layers.  The formation of the interlayer $X$-$X$ bonds has a dramatic effect on the DOS and position of the Fermi level.  In terms of the formal charge associated with the $X$ anions, the $X$-$X$ bond results in a [$X$-$X$]$^{-4}$ polyanion, as opposed to the separate $X^{-3}$ anions existing in the T phase \cite{Hoffman_1985}.  For the compounds $A$Co$_{2}$P$_{2}$ \mbox{($A=$ Ca, Sr, La, Ce, Pr, Nd, Sm, or Eu)}, which have the ThCr$_{2}$Si$_{2}$ structure, the distance between interlayer P cations correlates with the effective valence of the Co cations, and the valence of the Co cations affects both the structure of any magnetic order present at low temperatures and the value of the magnetic moment \cite{Reehuis_1998}.

The formation of bonds between $T_{2}X_{2}$ layers can explain the quenching of the Fe moment in the cT phase of CaFe$_{2}$As$_{2}$ \cite{Goldman_2008,Yildrim_2009}, the consequences of modifications to the Fermi-surface of BaFe$_{2}$As$_{2}$ due to structural distortions \cite{Kimber_2009}, as well as the magnetic phase diagrams for Sr$_{1-x}$Ca$_{x}$Co$_{2}$P$_{2}$ \cite{Jia_2009} and Ca$_{1-x}$Sr$_{x}$Co$_{2}$As$_{2}$ \cite{Ying_2013}.  For the case of CaFe$_{2}$As$_{2}$, the pressure-induced T-cT transition decreases the distance between Fe$_{2}$As$_{2}$ layers to a value consistent with the formation of an As-As bond, and band structure calculations for the cT phase show that a dramatically lower DOS at the Fermi level occurs along with a shift in the Fe $3d_{x^{2}-y^{2}}$ and  $3d_{xz+yz}$ bands to lower energies. The calculated generalized magnetic susceptibility indicates that the Fe $3d$ DOS at the Fermi level in the cT phase is insufficient to induce magnetic order \cite{Goldman_2009}.

We next compare our results to those for Ca$_{1-x}$Sr$_{x}$Co$_{2}$As$_{2}$, in which substitution of Sr for Ca is performed.  Though such doping is isoelectronic, Sr has a larger radius than Ca, which may induce steric effects.  For this series of compounds, several magnetic states occur: the expected A-type order for $x=0$, FM order with the ordered moments along the $c$ axis for $x=0.2$, another AFM order phase with A-type order and the moments lying in the $ab$ plane for $x=0.34$, and no magnetic order for $x\geq0.34$ \cite{Ying_2013}.  A deviation in $c(x)$ at $x\approx0.4$, from its smooth increase with increasing $x$, correlates with the disappearance of magnetic order and is assigned as the value at which the cT-T transition occurs \cite{Ying_2013}.  A study of the compounds Ca$_{1-x}$Sr$_{x}$Co$_{2}$P$_{2}$ gives a similar phase diagram, consisting of multiple magnetic states, and a T-cT phase transition at $x=0.5$.  In this case, the T-cT transition changes both the Co-Co and Co-P-Co bond lengths and also correlates with a transition from a nearly-ferromagnetic Fermi liquid to AFM order \cite{Jia_2009}.  For both compounds, transitions between the magnetically ordered ground states are also tied to changes in the lattice parameters \cite{Ying_2013, Jia_2009}.

\begin{figure}[]
	\centering
	\includegraphics[width=1.0\linewidth]{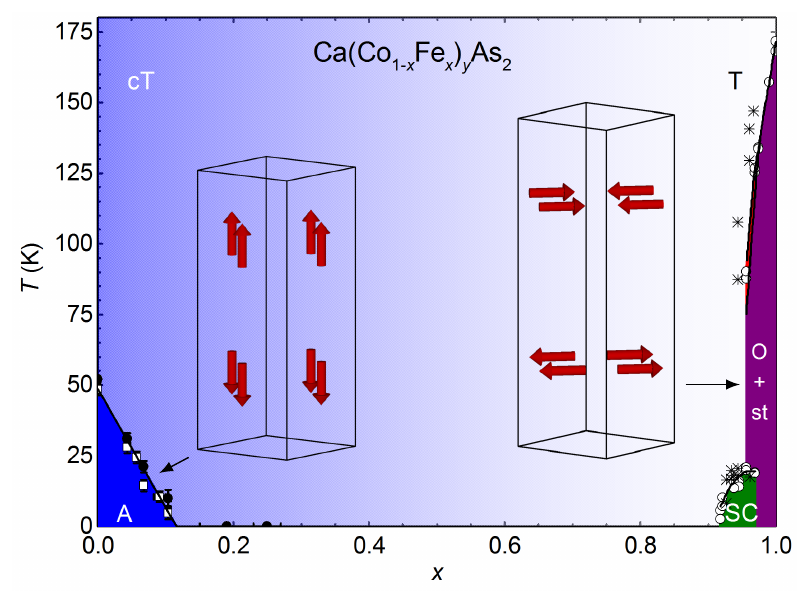}
	\caption{(Color online)  \label{Fig16} Phase diagram for Ca(Co$_{1-x}$Fe$_{x}$)$_{y}$As$_{2}$.  Points and lines for $x\geq0.91$ are from Fig.~$4$(a) in Ref.~\onlinecite{Hu_2012}, which includes data from Ref.~\onlinecite{Harnagea_2011}.  The A-type (A) and stripe-type (st) antiferromagnetic structures are illustrated with the collapsed-tetragonal (cT) and tetragonal (T) chemical unit cells, respectively, although the chemical unit cell is orthorhombic (O) for temperatures at which the stripe-type order occurs.  SC labels the superconducting region.}
\end{figure}

Figure~\ref{Fig16} shows the phase diagram for Ca(Co$_{1-x}$Fe$_{x}$)$_{y}$As$_{2}$ determined from this work and from data given in Refs.~\onlinecite{Harnagea_2011} and~\onlinecite{Hu_2012}.  In contrast to the results for  Ca$_{1-x}$Sr$_{x}$Co$_{2}$As$_{2}$ and Ca$_{1-x}$Sr$_{x}$Co$_{2}$P$_{2}$, which show that substitution of Sr for Ca causes multiple magnetic ground states and a clear cT-T transition accompanied by the suppression of magnetic order, our results show a monotonic suppression of A-type AFM order and a smooth crossover from the cT to the T phases.  This suggests that hole doping CaCo$_{1.86}$As$_{2}$ has a less dramatic effect than any steric effects due to doping Sr for Ca, and that it suppresses the A-type magnetic order in the absence of sharp changes to the lattice parameters or an abrupt cT-T phase transition.  

In addition to the effects of hole doping, the consequences of decreasing the vacancy of the Co site with increasing $x$ need to be considered.  This is especially important in light of the facts that samples synthesized using CoAs self flux have been reported as being either almost stoichiometric \cite{Ying_2013} or possessing  $5\%$ vacancy of the Co site \cite{Zhang_2015}.  Results from previous electronic structure calculations for CaCo$_{1.88}$As$_{2}$ and CaCo$_{2}$As$_{2}$ show that the stoichiometric compound has a $21.2\%$ higher total energy for the A-type AFM ground state, with a $7.4\%$ decrease of the Co DOS and an $8\%$ decrease in the total DOS at $E_{\textrm{F}}$ \cite{Anand_2014}.  They also indicate that the total energies of the A-type AFM and FM states are similar for both compositions, which points to competition between the two states, regardless of the presence of a $6\%$ vacancy.  Based on these results and the posited A-type AFM order for CaCo$_{2}$As$_{2}$, albeit occurring with a higher $T_{\textrm{N}}$ \cite{Cheng_2012, Ying_2013, Zhang_2015}, it is difficult to conclude what if any effect finite vacancy of the Co site has on the AFM structure of the ground state.  Nevertheless, the precise effects on the crossover from the cT to T structure warrants further studies, which, ideally, should be preformed by varying the amount of vacancy of the Co site without chemical substitution or charge doping.  Lastly, we point out that the larger uncertainty in the measured values of $y$ for the $x=0.67$ and $0.89$ samples indicates that the amount of disorder within these compounds  is larger than that for other values of $x$.  Such disorder may obscure the observation of a sharp phase transition, potentially leading to the experimentally observed crossover from the cT to T phase.

Finally, we compare the suppression of the A-type AFM order in Ca(Co$_{1-x}$Fe$_{x}$)$_{y}$As$_{2}$ with the sensitivity of Ca(Fe$_{1-x}$Co$_{x}$)$_{2}$As$_{2}$ to low levels of strain.  For FeAs-flux grown  Ca(Fe$_{1-x}$Co$_{x}$)$_{2}$As$_{2}$, single crystals annealed and then quenched at temperatures between $350\leq T<960$\,\degree C exhibit the low-temperature AFM and orthorhombic, and superconducting phases seen for samples grown using Sn flux, whereas as-grown crystals quenched from $960$\,\degree C enter the nonmagnetic cT phase upon cooling at ambient pressure \cite{Ran_2012, Ran_2011}.  Additionally, the first-order structural phase transition between the high-temperature T and low-temperature cT phases becomes less severe, or rather more continuous, with increasing Co concentration, and it is suggested that a critical end point exists, past which a continuous thermal contraction occurs rather than a dramatic first-order transition \cite{Ran_2012}.  The occurrence of a continuous T-cT phase transition with increasing Co concentration appears qualitatively similar to the smooth evolution with increasing $x$ from the cT to T phase for Ca(Co$_{1-x}$Fe$_{x}$)$_{y}$As$_{2}$.
    
\section{CONCLUSION }
We have shown that A-type AFM order exists in Ca(Co$_{1-x}$Fe$_{x}$)$_{y}$As$_{2}$ for $0\leq x < 0.12(1)$ with the moments lying along the $c$ axis.  We determine an ordered moment of  $\mu=0.43(5)~\mu_{\textrm{B}}/$Co at $T=4$~K for $x=0$, which agrees with previous estimates \cite{Anand_2014,Quirinale_2013}, and find that both $T_\textrm{N}$ and $\mu$ decrease with increasing $x$ with rates of $\frac{dT_{\textrm{N}}}{dx}=-418(27)$~K and $\frac{d\mu}{dx}=-3.24(5)\mu_{\textrm{B}}/$Co, respectively.  In addition, our neutron diffraction experiments find no evidence for the development of stripe-type AFM order for $x$ up to at least $0.25$ nor the development of FM order up to at least $x=0.104$.  X-ray diffraction data show a smooth evolution with increasing $x$ from the cT phase of  CaCo$_{1.86}$As$_{2}$  to the T phase of CaFe$_{2}$As$_{2}$.  Our results suggest that hole doping suppresses the A-type magnetic order without the occurrence of an abrupt cT-T phase transition, nor any other sharp changes to the lattice parameters, but do not fully address whether a finite amount of vacancy of the Co site or disorder smear out an inherently sharp cT-T transition.  To answer this question, systematic studies on the effects of partial vacancy of the magnetic site and chemical and structural disorder are necessary, and, in general, should lend more insight into the coupling between lattice, electronic, and magnetic degrees of freedom in CaCo$_{1.86}$As$_{2}$ and related compounds.
\\
\begin{acknowledgments}
We are grateful for assistance from D.\ Robinson with performing the high-energy x-ray diffraction experiments.  Work at the Ames Laboratory was supported by the U.\,S.\ Department of Energy (DOE), Basic Energy Sciences, Division of Materials Sciences \& Engineering, under Contract No.\ DE-AC$02$-$07$CH$11358$.  A portion of this research used resources at the High Flux Isotope Reactor, a U.\,S.\ DOE Office of Science User Facility operated by the Oak Ridge National Laboratory.  This research used resources of the Advanced Photon Source, a U.\,S.\ DOE Office of Science User Facility operated for the U.\,S.\ DOE Office of Science by Argonne National Laboratory under Contract No.\ DE-AC$02$-$06$CH$11357$.
\end{acknowledgments}

\end{document}